\documentclass[seceq]{ptptex}
\usepackage{graphicx}
\notypesetlogo  

\newcommand{\nc}{\newcommand}		
\newcommand{\renc}{\renewcommand}	
\nc{\vc}[1]	{\mbox{\boldmath $#1$}}	
\nc{\bs}[1]	{\boldsymbol{#1}}
\nc{\del}       {\partial}              
\nc{\bra}       {\langle}               
\nc{\ket}       {\rangle}               
\nc{\bras}[1]   {\langle #1|}           
\nc{\kets}[1]   {|#1\rangle}            
\nc{\mapleft}[1]{			
 \smash{\mathop{\,			%
  \hbox to 1.5cm{\rightarrowfill}\, }\limits_{#1}}}
  
\def\PPNP#1#2#3{ {{\rm Prog. Part. Nucl. Phys.}}\ \textbf{#1} (#2), #3}  

\begin{document}

\markboth{T. Myo, H. Toki and K. Ikeda}{   
Tensor-Optimized Shell Model with Bare $NN$ Interaction for $^4$He}

\title{Tensor-Optimized Shell Model \\ with Bare Nucleon-Nucleon Interaction for $^4$He}

\author{
Takayuki \textsc{Myo},$^{1,2,}$\footnote{\noindent E-mail: myo@ge.oit.ac.jp}
Hiroshi \textsc{Toki},$^{1,}$\footnote{\noindent E-mail: toki@rcnp.osaka-u.ac.jp}
and
Kiyomi \textsc{Ikeda}$^{1,3,}$\footnote{\noindent E-mail: k-ikeda@postman.riken.go.jp}
}

\inst{
$^1$Research Center for Nuclear Physics (RCNP), Osaka University, Ibaraki 567-0047, Japan\\
$^2$General Education, Faculty of Engineering, Osaka Institute of Technology, Osaka 535-8585, Japan\\
$^3$RIKEN Nishina Center, Wako 351-0198, Japan
}

\recdate{November 14, 2008}

\abst{
The pion exchange between nucleons generates a strong tensor interaction, which provides a large attractive contribution for the binding energy of nucleus.  This non central tensor interaction is difficult to handle in the shell model framework, which hinders full understanding of nuclear structure.  
We develop the tensor-optimized shell model (TOSM) for the strong tensor interaction and now we are able to use bare nucleon-nucleon interaction with the help of the unitary correlation operator method (UCOM) for the short-range hard core.
We adopt the nucleon-nucleon interaction, AV8$^\prime$, and calculate explicitly the ground state of $^4$He and make a detailed comparison with rigorous few-body model calculations.   We show a large amount of success of the tensor-optimized shell model
with bare nucleon-nucleon interaction for $^4$He.}

\maketitle

\newcommand{\beq}{\begin{eqnarray}}
\newcommand{\eeq}{\end{eqnarray}}
\newcommand{\vs}{\vspace{-0.275cm}}

\section{Introduction}\label{sec:intro}

It is important to develop a theoretical framework to calculate nuclear structure with many nucleons using the realistic nucleon-nucleon interaction, which is obtained from two nucleon scattering. Recently, it has become possible to calculate nuclei up to a mass of approximately $A\sim 12$ \cite{pieper01,pieper02,pudliner97} using the realistic nucleon-nucleon interaction.  The method used for the calculation is the Green's function Monte-Carlo method (GFMC) with the use of relative nucleon coordinates.   This method introduces various correlation functions with many variational parameters in the nuclear wave function.  In GFMC, the nuclear structures and binding energies were successfully reproduced by including three-body interaction. One big surprise is the extremely large contribution of the one pion exchange interaction, which is about 70 $\sim$ 80\% of the entire nucleon-nucleon interaction.  In principle, they can extend this method to calculate heavier nuclei.  It is, however, extremely time-consuming even with the present computer power.  Hence, it is strongly desired to develop a new method of calculating nuclei with large nucleon numbers using the nucleon-nucleon interaction.

The nucleon-nucleon interaction has distinctive features, namely there exist strong tensor interaction at intermediate distance caused by pion exchange and strong short-range repulsive interaction at a short distance caused by quark dynamics.  Although these two interactions have totally different characteristics, it is customary to adopt the Brueckner Hartree-Fock theory to integrate out the high-momentum components on the same footing and use the resulting $G$-matrix as an effective interaction in the shell model.  In this way, we lose information on the tensor correlation and a short-range correlation in the shell model wave function.  Hence, we search for a powerful method of treating treat explicitly both the tensor interaction and the short-range interaction to study of not only light nuclei but also medium and heavy nuclei.

There have been two important developments for this purpose.  One is the finding that the tensor interaction is of intermediate range, and hence, we can express the tensor correlation in a reasonable shell model space\cite{myo06,myo07}.  We call this method Tensor-Optimized Shell Model (TOSM), wherein the nuclear wave function is written in terms of the standard shell model state and a sufficient amount of two-particle two-hole ($2p2h$) states.  This TOSM formalism is based on the success of the parity and charge projection in the treatment of the pion exchange interaction\cite{sugimoto04,ogawa06}.  We have shown that the tensor interaction could be treated properly by taking a reasonable amount of multipoles ($l \leq 5$) in the $2p2h$ wave functions with the optimization of the radial parts of the particle states. The other is the Unitary Correlation Operator Method (UCOM) for the treatment of the short-range correlation\cite{feldmeier98,neff03}.  The short-range repulsive interaction is of very short-range and it is suited to treat the short-range correlation using unitary transformation and take the approximation to use only up to the two-body operators.  This approximation is justified because the volume associated with the short-range correlation is extremely small, where more than three nucleons rarely enter the small volume.  This is not the case for the tensor correlation, since the tensor interaction is of intermediate and long range as discussed by Neff and Feldmeier\cite{neff03}.

Our idea is to combine these two methods, TOSM and UCOM, to develop a theoretical framework that can describe medium and heavy nuclei beyond the light nuclei using the realistic nucleon-nucleon interaction.  
We can use the TOSM for the strong tensor interaction utilizing the intermediate nature caused by finite angular momentum of the relative wave function and the UCOM for the strong short-range interaction utilizing the short-range nature.  
We use completely different methods for these two distinctive characteristics of the nucleon-nucleon interaction.  
After demonstrating its power, we hope to apply the newly developed method, which we call TOSCOM, to many nuclei.  
Using TOSCOM, we aim to understand the roles of the tensor and short-range correlations in nuclei using bare interaction.
As a good start, we would like to apply TOSCOM to $^4$He.  
Hence, there are two purposes of this study.  
One is to see how this method works for the treatment of the bare nucleon-nucleon interaction.  
The other is to compare the obtained results with rigorous calculations to check the accuracy of TOSCOM.
From this comparison, we can see how far we can describe the short-range and tensor correlations and to determine what we need to do for the further improvement of TOSCOM in order to solve the nucleus as precisely as possible.

There are several methods for the description of few-body systems.  These methods are compared each other for $^4$He using the AV8$^\prime$ nucleon-nucleon interaction \cite{pudliner97,kamada01}.  These methods are called the Faddeev-Yakubovsky equations (FY)\cite{kamada92}, the coupled-rearrangement-channel Gaussian-basis variational method (CRCVM)\cite{hiyama03}, the stochastic variational method(SVM)\cite{varga95}, the hyperspherical harmonic method (HH)\cite{kievsky97}, GFMC \cite{carlson88}, the no-core shell model (NCSM)\cite{navratil99}, the effective interaction hyperspherical harmonic method (EIHH)\cite{novoselsky94}.  Although the methods of numerical calculations are largely different, these methods provide essentially the same results for the $^4$He structure and also the amount of kinetic energy and components of the nucleon-nucleon interaction.  All the methods described in Ref.~\citen{kamada01} have a common feature where the wave functions are expressed in terms of the relative nucleon coordinates (Jacobi coordinate).  We shall call this method as described here the $T$-coordinate method, since the relative coordinate between two nucleons is the coordinate of the nucleon-nucleon interaction.   Hence, it is easy to describe the correlations between two nucleons by taking sufficient variational variables.  This $T$-coordinate method is advantageous for rigorous calculations when there exists sufficient computer power.

On the other hand, we would like to develop a theoretical framework to describe wave functions in terms of single-particle coordinates, which we call $V$-coordinate method.  This $V$-coordinate method can be used to describe nuclei with many nucleons relatively easier than the $T$-coordinate method.  Furthermore, we are able to describe the wave function on the basis of the shell model picture, and hence, it becomes easier to interpret the calculated results in the shell model sense.  The difficulty, on the other hand, is to express the correlations of the relative motion between two nucleons, which are caused by the short-range repulsive interaction and the tensor interaction in nucleon-nucleon interaction.  We overcome this problem by developing TOSCOM to describe the short-range and tensor correlations simultaneously.

We also mention the differences of TOSCOM from the other works based on the shell model.
Recently, NCSM result has been reported for $^{40}$Ca starting from the realistic interaction, 
and discussions are made for the convergence of the binding energy.\cite{roth07} 
In their scheme, transformed interactions such as $V_{{\rm low}k}$, using UCOM and Lee-Suzuki technique\cite{navratil99}, are employed. 
In these interactions, high-momentum components of the short-range and tensor correlations are renormalized. 
Thus, the truncated shell model space calculation is applicable to discuss the binding energy.
On the other hand, in TOSCOM, the truncation of the model space is not introduced.
For the tensor interaction, we can directly use the bare tensor interaction to evaluate the matrix elements
and discuss the characteristics of the tensor interaction in the nuclear structure explicitly. 
For the short-range part, we employ UCOM, in which we can also obtain the explicit wave function including the short-range correlation through the UCOM transformation.

Otsuka et al. investigated the role of the 'tensor interaction', in particular, for neutron-rich nuclei.\cite{otsuka06,brown06}
They calculated the tensor interaction matrix elements mainly arising from the exchange term of the tensor interaction in the Hartree-Fock scheme.
On the other hand, we go beyond the Hartree-Fock approximation and treat the dominant part of the tensor interaction by taking the $2p2h$ wave functions up to all orders for the particle states.
In particular, the $0p0h$-$2p2h$ coupling of the tensor interaction is essential for describing the dominant part of the tensor interaction, which provides about half of the interaction matrix element.
In TOSCOM, this $2p2h$ contribution is explicitly included, and we focus on the strong tensor correlation represented by the $2p2h$ wave functions.

This paper is arranged as follows.  In \S \ref{sec:TOSM} and \S \ref{sec:UCOM}, we describe the TOSM + UCOM (TOSCOM) in detail.  
We shall apply TOSCOM to $^4$He in \S \ref{sec:result} and compare the numerical results with those of the few-body methods.  
In \S \ref{sec:conclusion}, this study is summarized and perspectives are described.

\section{Tensor-optimized shell model}\label{sec:TOSM}

We shall begin with many-body Hamiltonian,
\beq
H=\sum_i T_i - T_{\rm cm} + \sum_{i<j} V_{ij}
\label{eq:Ham}
\eeq
with
\begin{eqnarray}
    V_{ij}
&=& v_{ij}^C + v_{ij}^{T} + v_{ij}^{LS} + v_{ij}^{Clmb} .
\end{eqnarray}
Here, $T_i$ is the kinetic energy of all the nucleons with $T_{cm}$ being the center of mass kinetic energy.  
We take the bare nucleon-nucleon interaction for $V_{ij}$ such as the AV8$^\prime$ consisting of central ($v^C_{ij}$), tensor ($v^T_{ij}$) and spin-orbit ($v^{LS}_{ij}$) terms.  The $v_{ij}^{Clmb}$ is the Coulomb term. We describe the many-body system with many-body wave function, $\Psi$, by solving the equation $H \Psi=E \Psi$. In TOSCOM, we take the $V$-coordinates to express $\Psi$.
We explain here the typical features of nucleon-nucleon interaction, tensor interaction and short-range repulsion, and how we treat them in TOSCOM. 
The pion exchange interaction is a long- and intermediate-range interaction, which contains strong tensor interaction.  
This tensor interaction has a large strength in the intermediate range\cite{myo06}.  
Hence, we hope to describe the tensor correlation in terms of a reasonable amount of multipoles of single-particle states in the $2p2h$ states with high-momentum component.  
In fact, we have shown that the tensor correlation is expressed in terms of multipoles up to $l \leq 5$ for $^4$He \cite{myo06}.  
On the other hand, there exists a strong repulsive interaction in the short-range part of the nucleon-nucleon interaction.  This is the other difficulty to be considered for the nuclear many-body problem. 
For this problem, Feldmeier et al. have demonstrated that UCOM can be used to treat the short-range correlation\cite{feldmeier98,neff03}.

We begin with TOSM and write the case of $^4$He explicitly as an example.  The wave function, $\Psi$, is written as
\beq
\Psi=C_0 \kets 0 + \sum_p C_p \kets {2p2h}_p .
\eeq
Here, the wave function $\kets {0}$ is a shell model wave function and $\kets {(0s)^4}$ for $^4$He.  $\kets {2p2h}$ represents a $2p2h$ state with various ranges for the radial wave functions of particle states. We can write $\kets {2p2h}$ as
\beq
    \kets {2p2h}_p
&=& \kets {[ [\psi^{n_1}_{\alpha_1}(\vec x_1)\psi^{n_2}_{\alpha_2}(\vec x_2)]^J
    \otimes [\tilde\psi^{n_3}_{\alpha_3}(\vec x_1)\tilde\psi^{n_4}_{\alpha_4}(\vec x_2)]^J]^0}_A \ .
\label{eq:2p2h}
\eeq
The suffix $A$ of the wave function indicates anti-symmetrization of the wave functions.  Here, $p$ denotes representable quantum number of $2p2h$ states, which are expressed with particle (hole) wave functions $\psi^{n}_{\alpha}$ ($\tilde\psi^{n}_{\alpha}$).
The index, $n$, is to distinguish the different radial components of the single-particle wave function, $\psi$.
The index, $\alpha$, is a set of three quantum numbers, $l$, $j$ and $t_z$, to distinguish the single-particle orbits, 
where $l$ and $j$ are the orbital and total angular momenta of the single-particle states, respectively, and $t_z$ is the projection of the nucleon isospin.  
The normalization factors of the two particle states are included in the wave functions given in Eq.~(\ref{eq:2p2h}).
For $^4$He, the coupled spin, $J$, of two nucleons is $J=0$ or $J=1$. We omit writing the coupled isospin, which should be either 0 or 1 depending on the value of $J$. We have used Gaussian functions for radial wave functions to express more effectively compressed radial wave functions\cite{myo06}.  
The shell model technique is used to calculate all the necessary matrix elements, which are expressed explicitly in the Appendix.

We explain the Gaussian expansion technique for single-particle orbits\cite{hiyama03,aoyama06}. 
Each Gaussian basis function has the form of a nodeless harmonic oscillator wave function (HOWF), except for $1s$ orbit.
When we superpose a sufficient number of Gaussian bases with appropriate length parameters, we can fully optimize the radial component of every orbit of every configuration with respect to the total Hamiltonian in Eq.~(\ref{eq:Ham}).
We construct the following ortho-normalized single-particle wave function $\psi^n_{\alpha}$ with a linear combination of Gaussian bases $\{\phi_\alpha\}$ with length parameter $b_{\alpha,m}$.
\begin{eqnarray}
        \psi^n_{\alpha}(\vc{r})
&=&     \sum_{m=1}^{N_\alpha} d^n_{\alpha,m}\ \phi_{\alpha}(\vc{r},b_{\alpha,m})
        \qquad
        {\rm for}~~n~=~1,\cdots,N_\alpha
        \label{eq:Gauss1}
\end{eqnarray}
Here, $N_\alpha$ is the number of basis functions for $\alpha$, and $m$ is an index that distinguishes the bases with different values of $b_{\alpha,m}$.
The explicit form of the Gaussian basis function is expressed as
\begin{eqnarray}
        \phi_{\alpha}(\vc{r},b_{\alpha,m})
&=&     N_l(b_{\alpha,m})\ r^l\ e^{-(r/b_{\alpha,m})^2/2}\ [Y_{l}(\hat{\bs{r}}),\chi^\sigma_{1/2}]_j \chi_{t_z},
        \label{eq:Gauss2}
        \\
        N_l(b_{\alpha,m})
&=&     \left[  \frac{2\ b_{\alpha,m}^{-(2l+3)} }{ \Gamma(l+3/2)}\right]^{\frac12}.
\end{eqnarray}
The coefficients $\{d^n_{\alpha,m}\}$ are determined by solving the eigenvalue problem for the norm matrix of the non orthogonal Gaussian basis set in Eq.~(\ref{eq:Gauss2}) with the dimension $N_\alpha$.
Following this procedure, we obtain new single-particle wave functions $\{\psi^n_{\alpha}\}$ using Eq.~(\ref{eq:Gauss1}).

We choose the Gaussian bases for the particle states to be orthogonal to the occupied single-particle states, which is $0s_{1/2}$ in the $^4$He case. For $0s_{1/2}$ states, we employ one Gaussian basis function, namely, HOWF with length $b_{0s_{1/2},m=1}=b_{0s}$. 
For $1s_{1/2}$ states, we introduce an extended $1s$ basis function orthogonal to the $0s_{1/2}$ states 
and possessing a length parameter $b_{1s,m}$ that differs from $b_{0s}$ \cite{myo06}. 
In the extended $1s$ basis functions, we change the polynomial part from the usual $1s$ basis states to satisfy the conditions of the normalization and the orthogonality to the $0s$ state. 

Two-body matrix elements in the Hamiltonian are analytically calculated using the Gaussian bases\cite{hiyama03,aoyama06}, whose explicit forms are given in the Appendix for central, LS and tensor interactions, respectively.
In the numerical calculation, we prepare 9 Gaussian functions at most with parameters of various ranges to obtain a convergence of the energy.

Furthermore, we have to take care of the center-of-mass excitations.  For this purpose, we use the well-tested method of introducing a center-of-mass term in the many-body Hamiltonian\cite{otsuka07,lawson}.  
\begin{eqnarray}
       H_{\rm cm}
&=&    \lambda \ \left( \frac{\vc{P}_{\rm cm}^2}{2A\, m}+ \frac12\, A\, m\, \omega^2\, \vc{R}_{\rm cm}^2-\frac{3}{2}\hbar\omega \right),
       \\
        \vc{P}_{\rm cm}
&=&     \sum_{i=1}^A \vc{p}_i,\quad
        \vc{R}_{\rm cm}
\,=\,   \frac1A\, \sum_{i=1}^A \vc{r}_i,\quad
        \omega
\,=\,   \frac{\hbar}{m\, b_{0s}^2}
\end{eqnarray}
Here, $m$ and $A$ are the nucleon mass and the mass number, respectively, and
$b_{0s}$ is the length parameter of HOWF for the hole $0s$ state.
We take a sufficiently large coefficient, $\lambda$, to project out only the lowest HO state for the center-of-mass motion.
In the numerical calculation, the excitation of the spurious center-of-mass motion is suppressed to be less than 10 keV.

The variation of the energy expectation value with respect to 
the total wave function $\Psi(^{4}{\rm He})$ is given by
\begin{eqnarray}
\delta\frac{\bra\Psi|H|\Psi\ket}{\bra\Psi|\Psi\ket}&=&0\ ,
\end{eqnarray}
which leads to the following equations:
\begin{eqnarray}
    \frac{\del \bra\Psi| H - E |\Psi \ket} {\del b_{\alpha,m}}
&=& 0\ ,\quad
    \frac{\del \bra\Psi| H - E |\Psi \ket} {\del C_{p}}
=   0\ .
   \label{eq:vari}
\end{eqnarray}
Here, $E$ is a Lagrange multiplier corresponding to the total energy. The parameters $\{b_{\alpha,m}\}$ for the Gaussian bases appear in non linear forms in the energy expectation value. We solve two types of variational equations in the following steps. First, fixing all the length parameters $b_{\alpha,m}$, we solve the linear equation for $\{C_{p}\}$ as an eigenvalue problem for $H$ with partial waves up to $L_{\rm max}$. We thereby obtain the eigenvalue $E$, which is a function of $\{b_{\alpha,m}\}$. Next, we try to search various sets of the length parameters $\{b_{\alpha,m}\}$ to find the solution that minimizes the total energy. In this wave function, we can describe the spatial shrinkage with an appropriate radial form, which is important for the tensor correlation.\cite{myo06}

\section{Formulation of UCOM}\label{sec:UCOM}

\subsection{short-range correlation in UCOM}
We employ UCOM for the short-range correlation.  Feldmeier et al. worked out a unitary correlation operator in the form \cite{feldmeier98,neff03},
\beq
C&=&\exp\left(-i\sum_{i<j} g_{ij}\right)~=~\prod_{i<j}c_{ij} 
\eeq
with $c_{ij}= \exp(-i\ g_{ij})$. Here, $i$ and $j$ are the indices to distinguish particles.  
Here, the two-body operator, $g_{ij}$, is a Hermite operator, and hence, $C$ is a unitary operator.  We express the full wave function, $\Psi$, in terms of less sophisticated wave function, $\Phi$, as $\Psi=C\Phi$.  Hence, the Schr\"oedinger equation, $H\Psi=E\Psi$ becomes $\hat{H}\Phi=E\Phi$, where $\hat{H}=C^\dagger H C$.  If we choose properly the unitary correlator, $C$, we are able to solve more easily the Schr\"oedinger equation.  Moreover, once we obtain $\Phi$, we can then obtain the full wave function, $\Psi$, by the unitary transformation, $\Psi=C\Phi$.  Since $C$ is expressed with a two-body operator in the exponential, it is a many-body operator.  In the case of the short-range correlation, we are able to truncate modified operators at the level of two-body operators\cite{feldmeier98}.

In the actual calculation of UCOM, we define the operator $g_{ij}$ as
\begin{equation}
g_{ij}=	\frac12 \left\{ p_{r,ij} s(r_{ij})+s(r_{ij})p_{r,ij}\right\},
\end{equation}
where the momentum $p_{r,ij}$ is the radial component of the relative momentum, which is conjugate to the relative coordinate $r_{ij}$.  $s(r_{ij})$ is the amount of the shift of the relative wave function at the relative coordinate, $r_{ij}$, for each nucleon pair. 
Hereafter, we omit the indices $i$ and $j$ for simplicity. We also introduce $R_+(r)$ as
\begin{equation}
        \int_r^{R_+(r)}\frac{d\xi}{s(\xi)} =     1 ,
\end{equation}
which leads to the following relation,
\begin{equation}
	\frac{dR_+(r)}{dr}
=     \frac{s \left( R_+(r) \right)}{s(r)} .
\end{equation}
In UCOM, we use $R_+(r)$ instead of $s(r)$ to use the UCOM prescription. $R_+(r)$ represents the correlation function to reduce the amplitude of the short-range part of the relative wave function in nuclei and can be determined for four spin-isospin channels independently.
The explicit form of the transformation of the operator for the relative motion is given as
\begin{eqnarray}
    	c^\dagger r c
&=&	R_+(r) , 
        \qquad
        c^\dagger p_r c
~=~     \frac{1}{\sqrt{R^\prime_+(r)}}  p_r \frac{1}{\sqrt{R^\prime_+(r)}} , 
        \qquad
        c^\dagger \vc{l} c
~=~     \vc{l} ,
        \\
        c^\dagger \vc{s} c
&=&     \vc{s} , 
        \qquad
        c^\dagger S_{12} c
~=~     S_{12} ,
        \qquad
        c^\dagger v(r)  c
~=~  v(R_+(r)) , 
\end{eqnarray}
where the operators $\vc{l}$, $\vc{s}$ and $S_{12}$ are the relative orbital angular momentum operator,
intrinsic spin operator and the tensor operator, respectively. $v(r)$ is the arbitrary function depending on $r$, such as potential.

In the calculation using UCOM, we parametrize $R_+$(r) in the same manner as proposed by Neff-Feldmeier and Roth et al.\cite{feldmeier98,neff03,roth06}.
We assume the following forms for even and odd channels, respectively.
\begin{eqnarray}
	R_+^{\rm even}(r)
&=&	r + \alpha \left(\frac{r}{\beta}\right)^\gamma \exp[-\exp(r/\beta)],
	\\
	R_+^{\rm odd}(r)
&=&	r + \alpha \left( 1- \exp(-r/\gamma) \right) \exp[-\exp(r/\beta)]
\end{eqnarray}
Here, $\alpha$, $\beta$, $\gamma$ are the variational parameters to optimize the function $R_+(r)$ and minimize the energy of the system.
They are independently determined for four channels of the spin-isospin pair.
In the actual procedure of the variation, once we fix the parameters included in $R_+(r)$, we solve the eigenvalue problem 
of the Hamiltonian using Eq.~(\ref{eq:vari}) and determine the configuration mixing of the shell model-type bases.
Next, we try to search various sets of the $R_+(r)$ parameters to minimize the obtained energy.

\subsection{Extension of UCOM  --S-wave UCOM--}\label{sec:SUCOM}

In the framework of UCOM, we introduce the UCOM function $R_+(r)$ 
for each spin-isospin channel and ignore the partial wave dependence of $R_+(r)$.
It is generally possible to introduce the partial wave dependence in UCOM and then
$R_+(r)$ functions are determined in each relative partial wave in the two-body matrix elements.
Here, we consider the specific case of this extension of UCOM by taking care of the characteristics of the short-range correlation.
One of the simplest cases of this extension is UCOM for only $s$-wave relative motion,
since all the other partial waves $l$ except for $s$-wave ($l=0$) have $r^l$ behavior near the origin, 
where the short-range hard core is extremely large.  
Hence, this $r^l$ behavior largely cuts down the effect of the short-range hard core.
However, only the $s$-wave function is finite at the origin, and the behavior in the origin is determined by the hard core dynamics.  
In fact, the method used by Feldmeier et al. is to determine the unitary operator to reproduce the short-range behavior of $s$-wave relative wave function.

When we incorporate $S$-wave UCOM ($S$-UCOM, hereafter) into TOSM, we extract the relative $s$-wave component in all the two-body matrix elements in TOSM 
using the $V$-type basis expanded by the Gaussian functions.
For numerical calculations, we prepare the completeness relation consisting of the $T$-type basis functions $\kets{T}$ as
\begin{eqnarray}
      1
&=&   \sum_i |T_i \ket \bra T_i|,
      \qquad
      |T_i\ket
~=~   |[[\psi^{\boldsymbol{r}}_{l}\psi^{\boldsymbol{R}}_L]_{L'}, \chi_{S}]_J\ \chi_T \ket ,
      \label{eq:T}
\end{eqnarray}
where the $T$-type basis is expanded by the two coordinates of the relative part $\vc{r}$ and the center of mass part
$\vc{R}$ of two nucleons, which are the set of Jacobi coordinate.
The orbital angular momenta of each coordinate, $\vc{r}$ and $\vc{R}$, are $l$ and $L$, respectively.
It is easy to prepare the $s$-wave relative part by considering $l$ as zero in the $T$-type basis.
We construct the above completeness relation of the $T$-type basis states by diagonalizing the norm matrix 
expanded by the finite number of Gaussian basis functions for two coordinates. 
In the actual calculation, we use 12 bases for each coordinate, with which convergence is achieved.

We calculate the matrix elements of the arbitrary two-body operator $\hat{O}$ including the $S$-UCOM correlator $C_s$
using the $V$-type basis with indices $\alpha$ and $\beta$.
Here, we insert the above $T$ type completeness relation in Eq.~(\ref{eq:T}) as
\begin{eqnarray}
      \bra V_\alpha | C_s^\dagger \hat{O} C_s | V_\beta \ket
&=&   \sum_{ij} \bra V_\alpha |T_i \ket \cdot 
      \bra T_i| C_s^\dagger \hat{O} C_s |T_j \ket \cdot  
      \bra T_j| V_\beta \ket.
\end{eqnarray}
The matrix element using $T$-type base, $\bra T_i| C_s^\dagger \hat{O} C_s |T_j \ket$, is calculated for the two-body kinetic part and central, tensor interactions.
For the kinetic part and the central interaction, the matrix elements conserve the relative angular momentum, and then we can easily calculate the matrix elements of the transformed operator $C_s^\dagger \hat{O} C_s$.
For the tensor interaction, the $sd$ coupling matrix elements are properly treated, in which $C_s$ is operated on only the $s$-wave part 
of the relative motion. In this case, the operator $C_s$ acts on the $s$-wave relative Gaussian basis function $\phi_{l=0}(r)$, which is transformed as
\begin{eqnarray}
      C_S\ \phi_{l=0}(r)
&=&  \frac{R_-(r)}{r}\ \sqrt{R^\prime_-(r)}\ \phi_{l=0}(R_-(r)),
\end{eqnarray}
where $R_-(r)$ is the inverse transformation of $R_+(r)$, namely, $R_-(R_+(r))=r$.
The matrix elements of the $T$-type basis function are calculated using the above transformed wave function. 
We also calculate the overlap between $V$ type and $T$-type bases using Gaussian basis functions, whose explicit form is given in the Appendix.

\section{Numerical results of TOSCOM}\label{sec:result}

In this section, we show our numerical results for $^4$He using AV8$^\prime$ potential, which consists of central, LS and tensor terms
and is used in the rigorous calculation given by Kamada et al. where the Coulomb term is ignored\cite{kamada01}.

\subsection{Optimization of $R_+(r)$}
\begin{figure}[th]
\centering
\includegraphics[width=8cm,clip]{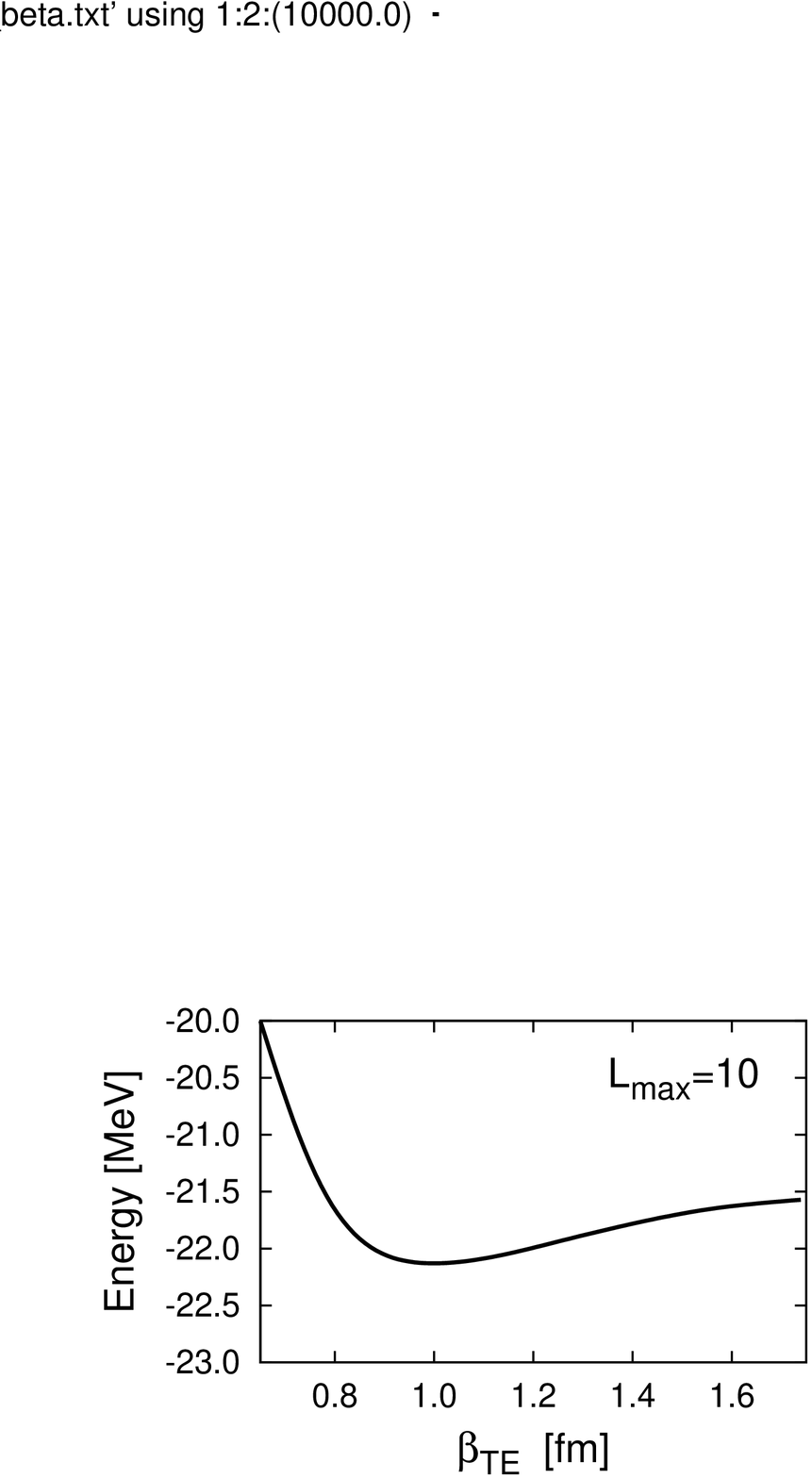}
\caption{Energy surface of $^4$He with respect to $\beta$ of triplet even channel with $L_{\rm max}$ being 10.}
\label{beta}
\end{figure}

\begin{figure}[th]
\centering
\includegraphics[width=6.0cm,clip]{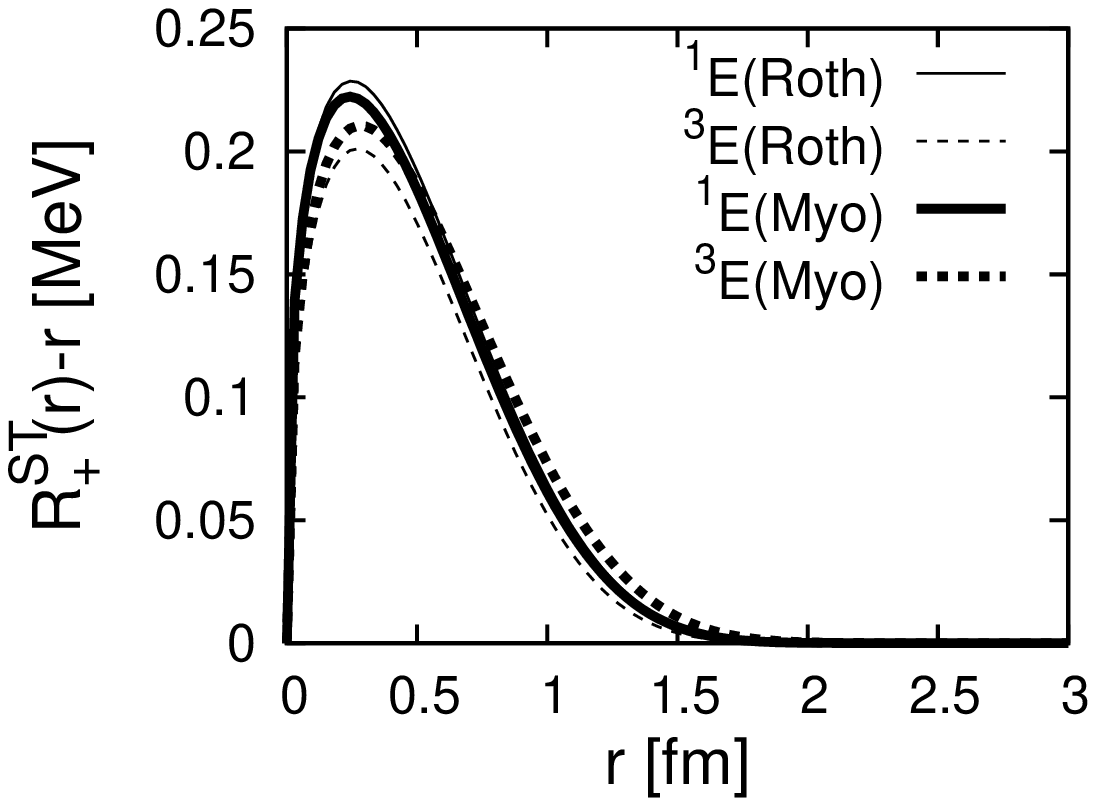}
\hspace*{0.75cm}
\includegraphics[width=6.0cm,clip]{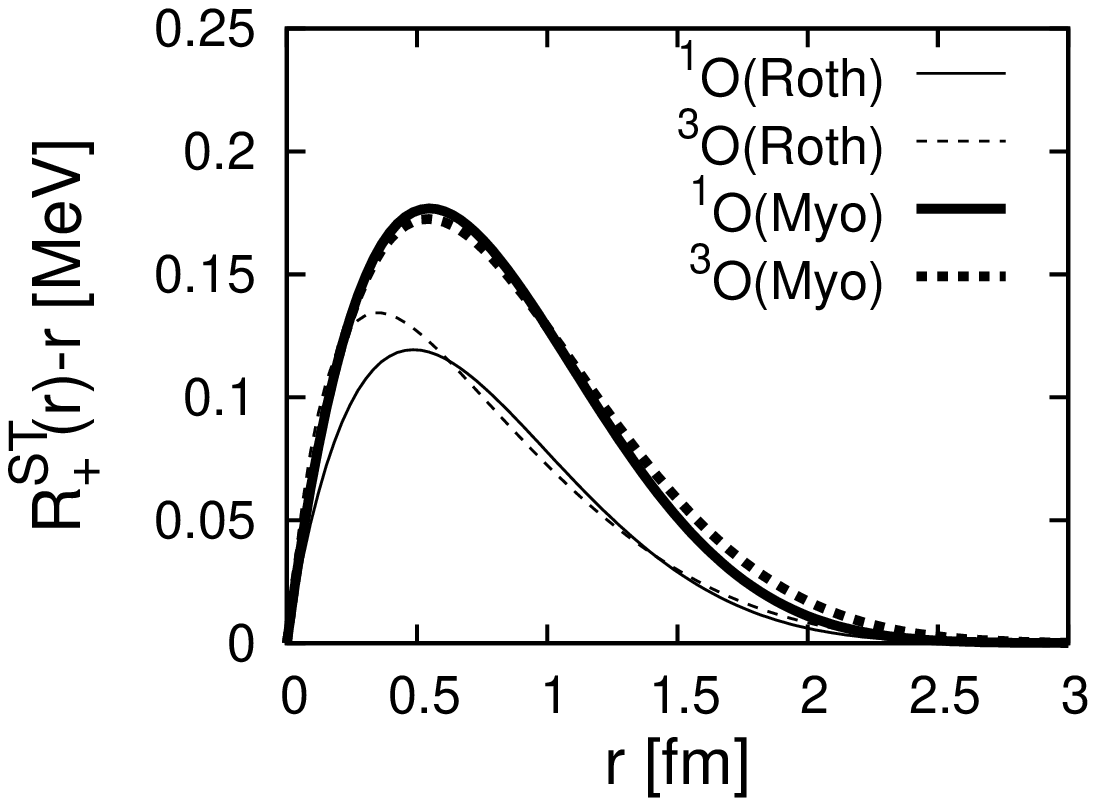}
\caption{Short-range correlation functions, $R_+(r)$, for UCOM in even and odd channels.  The thin curves are for the $R_+(r)$ function of Roth et al. and the thick curves are for that of this work.}
\label{R+}
\end{figure}

First, we determine the UCOM functions $R_+(r)$ for the calculation of TOSCOM.
In UCOM, we optimize $R_+(r)$ function by changing the three parameters of $\alpha$, $\beta$ and $\gamma$ to search for the energy minimum in TOSCOM. 
In Table \ref{tab:R+2}, the optimized three parameters in $S$-UCOM are listed.
The demonstration of the calculated result to search for the energy minimum is shown in Fig. \ref{beta}.
In Fig. \ref{beta}, the total energy is plotted as a function of the range parameter, $\beta$ of $R_+(r)$ function of the triplet even channel in the case of $L_{\rm max}$ being 10.
We have already minimized $\alpha$ and $\gamma$ for each $\beta$ in this calculation.  
We see clearly a desired behavior, where the energy has a minimum as a function of $\beta$.  
Hence, we fix $\beta$ that provides the lowest energy.
In Fig. \ref{R+}, $R_+(r)$ functions used in the present study are plotted in comparison with the case in Ref.~\citen{roth06}.
For the odd channel, in accordance with the discussion in Refs. \citen{neff03} and \citen{roth06}, we cannot find the optimum value of $R_+(r)$
in the two-body cluster approximation of the UCOM transformation for the Hamiltonian.
Hence, we decide to fix the range of $R_+(r)$, namely, $\beta$ as the same one adopted in Ref.~\citen{roth06} 
and optimize $\alpha$ and $\gamma$, while the variation of $R_+(r)$ for the odd channel does not have significant effects on the energy and other properties of $^4$He in comparison with the original case\cite{neff03,roth06}. 
Essentially, two types of parameter set of $R_+(r)$ in the present study and Ref.~\citen{roth06} 
give the similar form of $R_+(r)$ for even channels, in which we omitted the correlation function for the even channels except for $s$-waves. This result indicates that the correlation functions for the short-range repulsion are uniquely determined for each channel.

\begin{table}[t]
\begin{center}
\caption{Optimized parameters in $R_+(r)$ in TOSCOM for four channels in fm in the present work.}
\label{tab:R+2} 
\begin{tabular}{c|cccc}
              &  $\alpha$ &  $\beta$  &  $\gamma$ \\
\hline\hline
singlet even  &  1.32  &  0.88  &  0.36 \\
triplet even  &  1.33  &  0.93  &  0.41 \\
singlet odd   &  1.57  &  1.26  &  0.73 \\
triplet odd   &  1.18  &  1.39  &  0.53 \\
\hline
\end{tabular}
\end{center}
\end{table}

\begin{figure}[t]
\centering
\includegraphics[width=10cm,clip]{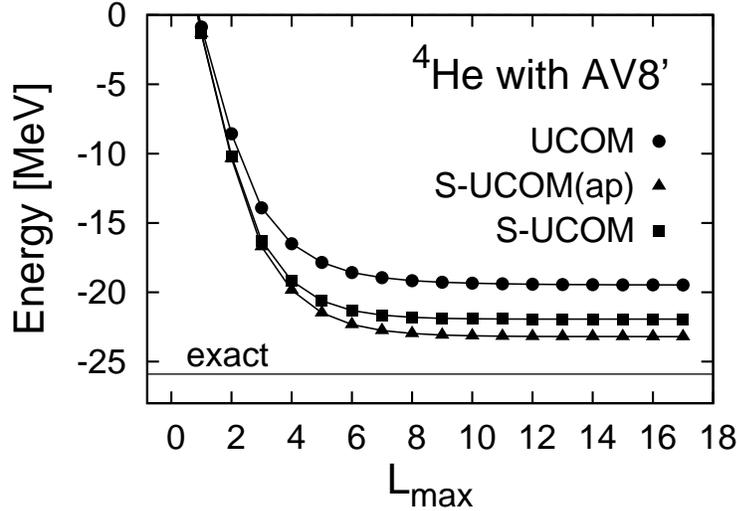}
\caption{Energy of $^4$He in TOSCOM as a function of the maximum angular momentum, $L_{\rm max}$.  
The circles are the results where UCOM is used for all the partial waves. 
The triangles ($S$-UCOM(ap)) are the results using $S$-UCOM with the approximation neglecting the UCOM effect in the tensor interaction matrix elements.
The squares are the results using $S$-UCOM accurately.}
\label{energy}
\end{figure}

\subsection{TOSCOM results for $^4$He}
Next, we show the calculated results of the energy of $^4$He as a function of $L_{\rm max}$ in Fig.~\ref{energy}.
We shall then compare the obtained results with the benchmark calculation given in Ref.~\citen{kamada01}.  
To start with, we show the ordinary UCOM case where UCOM is used for all the partial waves.
The calculated results of the energy are indicated in Fig. \ref{energy} by circles as a function of the maximum angular momentum, $L_{\rm max}$.  
The results show good convergence to reach $-19$ MeV, while the exact value of the few-body calculations is approximately $-26$ MeV as indicated in Fig. \ref{energy}.
We would like to point out that we can calculate the binding energy directly using the nucleon-nucleon interaction in TOSCOM.  
However, the binding energy is small.  The tensor interaction matrix element is approximately $-50$ MeV. 
On the other hand, in the previous study\cite{myo06}, we obtained approximately $-60$ MeV for the tensor interaction matrix element to check the validity of TOSM, 
when we used $G$-matrix for the central interaction to renormalize the short-range repulsion and retained the bare tensor interaction of AV8$^\prime$ in our previous calculation.
This fact indicates that the treatment of the short-range repulsive interaction is interfering with the contribution of the tensor interaction.  
This is due to a large removal of the short-range part of the relative wave functions in UCOM, in particular, in the $d$-wave part of the $sd$ coupling of the tensor interaction matrix element, 
where the tensor interaction possesses some amount of strength.
We have also calculated the contributions beyond the $2p2h$ configurations in TOSM such as $3p3h$ and $4p4h$ configurations.
When we include the $4p4h$ configurations within the $p$-shell,  their contribution to the binding energy is approximately 50 keV.
This fact denotes that these more complicated wave functions contribute very little in the total $^4$He wave function. 

We have decided to restrict the use of UCOM to the relative $s$-wave only for the even channel ($S$-UCOM), where the treatment of the short-range repulsion is absolutely necessary. 
In other partial waves, we have the centrifugal potential that cuts out the short-range part from the wave functions of the higher partial waves.  
In this case, we can use the modified interaction and the kinetic energy only for the relative $s$-wave component in the even channels.  
Since the use of the UCOM for the odd partial wave is slightly better, we use the UCOM for all odd partial waves.  
As a starting calculation, we have neglected the $S$-UCOM correlation in the calculation of the tensor interaction matrix elements.  
The results in this case ($S$-UCOM(ap)) are indicated also in Fig.~\ref{energy} by triangles.  
The energy converges to $-24$ MeV, which is now very close to the exact one as shown in Fig. \ref{energy}. 
In this case, the tensor interaction matrix element is $-61$ MeV, which becomes close to the exact value of $-68$ MeV.  
This improvement mainly comes from the increase in the $sd$ coupling of the tensor interaction matrix element,
however, this calculation is still not yet perfect.
We have to treat the effects of the short-range repulsion on the tensor interaction matrix element.
Hence, we have worked out the formulation to treat the rigorous $s$-wave function with the effect of the short-range repulsion 
for the calculation of the tensor interaction matrix elements as explained in the previous section.

\begin{figure}[t]
\centering
\includegraphics[width=10cm,clip]{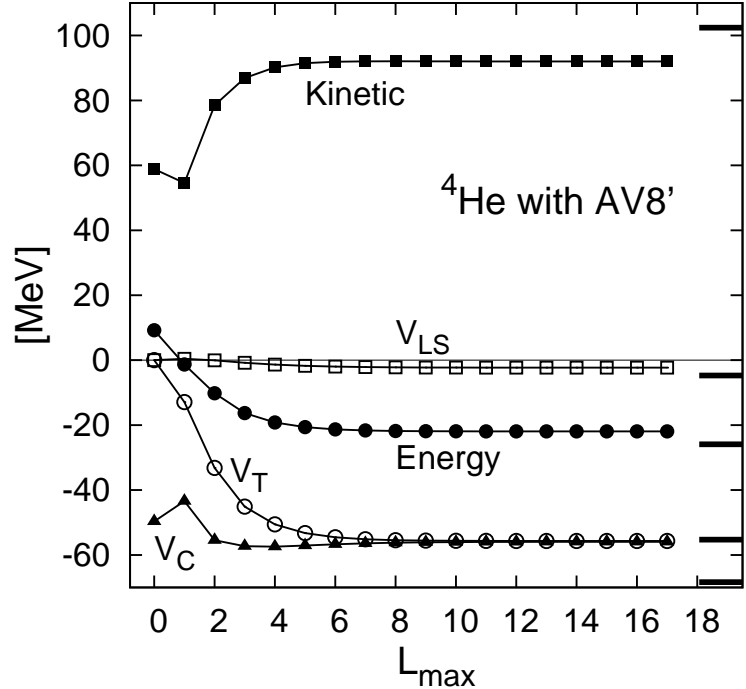}
\caption{Matrix elements of the central interaction ($V_C$), tensor interaction ($V_T$) and the spin-orbit interaction ($V_{LS}$) together with the kinetic energy (Kinetic) and total energy (Energy) in the Hamiltonian for $^4$He as function of $L_{\rm max}$. We observe good convergence for all the matrix elements.  These values are compared with the benchmark results of Ref.~\citen{kamada01}, which are indicated by the thick short solid lines on the right-hand side of the figure.}
\label{H_comp}
\end{figure}

The numerical calculation is quite involved in the $S$-UCOM case, since the $s$-wave relative wave function with the effect of short-range repulsion should be used for the tensor interaction matrix element.  
We show the calculated results for the total energy by the squares in Fig. \ref{energy}.  
We see quite a satisfactory result for the total energy, which is approximately $-22$ MeV. 
We show now all the components of the energy for $^4$He in Fig. \ref{H_comp}.
All the energy components show the saturation behavior as function of $L_{\rm max}$.  
In the tensor component, the saturation is obtained at around $L_{\rm max}$ being $8$.  
For the other components, their saturation points are seen at the similar $L_{\rm max}$.  
A very interesting feature is the kinetic energy, which goes up to a large value as the tensor interaction matrix element becomes large. 
As for the comparison with the rigorous calculation, we see that $V_c$ satisfies the rigorous value, which is approximately $-55$ MeV.  
On the other hand, the tensor interaction matrix element, $V_T$ converges to $-55$ MeV, while the rigorous one is $-68$ MeV.  
The kinetic energy is approximately $90$ MeV, while the rigorous one is $102$ MeV.  
The LS matrix element is also smaller than the rigorous value. 
As the net value, the total energy, $E$, is $-22$ MeV and the rigorous value is $-26$ MeV.   
A detailed comparison is performed in Table \ref{tab:benchmark}, in which the converged energies in TOSCOM are shown with the rigorous calculations. 
One of the possibilities for the lack of the energy in TOSCOM
is due to the separate treatment of the short-range and tensor correlations.
Although the dominant part of the tensor interaction is of intermediate and long range,
there may remain some small strength in the short-range part of the tensor interaction, which can couple with the short-range correlations. 
This effect can be included by extending the truncation of the UCOM transformation in the Hamiltonian to more than the two-body level.
Three-body term of the UCOM transformation is one of the possibilities to overcome the lack of energy in TOSCOM.\cite{feldmeier98}

\begin{table}[t]
\caption{Total energy, matrix elements of the Hamiltonian and radius of $^4$He compared with the benchmark results in Ref.~\citen{kamada01}.
Units are in MeV for the total energy and the matrix elements, and fm for the radius of $^4$He.}
\label{tab:benchmark} 
\begin{center}
\begin{tabular}{c|cccccc}
                           & Energy     &  Kinetic &  Central   & Tensor   & LS     & Radius \\
\hline\hline
Present(UCOM)              &  $-$19.46  &   88.64  &  $-56.81$ & $-50.05$ & $-1.24$ & 1.555 \\
Present($S$-UCOM)          &  $-$22.30  &   90.50  &  $-55.71$ & $-54.55$ & $-2.53$ & 1.546 \\
FY in Ref.\citen{kamada01} &  $-$25.94  &  102.39  &  $-55.26$ & $-68.35$ & $-4.72$ & 1.485 \\
\hline
\end{tabular}
\end{center}
\end{table}

We show the properties of the $^4$He wave functions obtained in the present calculation in Table \ref{prob}.
The three probabilities of the total intrinsic spin $S$ are shown in this table, 
in which the coupled value of the intrinsic spin of four nucleons is calculated without the UCOM transformation.
It is found that the $S=2$ component is larger than the $S=1$ case. 
This $S=2$ component is strongly caused by the tensor interaction, which can change the spin of two nucleon pairs by two.
\begin{table}[t]
\caption{Probabilities of the total intrinsic spin $S$ components in \%.}
\label{prob} 
\begin{center}
\begin{tabular}{cccc}
\hline
\hline
$S=0$   &  $S=1$    &  $S=2$   \\
\hline
89.41   &  2.47     &   8.12   \\
\hline
\end{tabular}
\end{center}
\end{table}
We show also the expectation values of the potentials for four channels in Table \ref{channel}.
The large contributions of the triplet even channel ($^3E$) are shown for the tensor and LS terms.
On the other hand, the odd channel contributions are very small.
\begin{table}[t]
\caption{Expectation values of the potentials for four channels.}
\label{channel} 
\begin{center}
\begin{tabular}{c|cccccc}
                          &  $^1E$  &  $^3E$     & $^1$O   & $^3$O   \\
\hline
\hline
Central                  & $-37.49$ &  $-19.13$  & $0.47$  & $0.45$  \\
Tensor                   &   --     &  $-54.03$  & --      & $-0.52$ \\
LS                       &   --     &  $-3.07$   & --      & $0.54$  \\ 
\hline
\end{tabular}
\end{center}
\end{table}

In Table~\ref{tab:mixing}, we list the mixing probabilities of the dominant configurations in $^4$He.
The subscripts 00 and 10 represent $J$ and $T$, the spin and isospin quantum numbers, respectively.
It is found that the $2p2h$ configurations with ($J$, $T$)=$(1,0)$ for the particle pair state are significantly mixed.
These spin and isospin are the same as those for the deuteron, and thus, this two-nucleon coupling can be understood as a deuteron-like correlation\cite{myo06}.

\begin{table}[t]
\caption{Mixing probabilities in the $^4$He ground state in \%.}
\label{tab:mixing} 
\begin{center}
\begin{tabular}{c|c}
\hline
\hline
$(0s)_{00}^4$                                  & 82.48  \\
$(0s)_{10}^{-2}(0p_{1/2})_{10}^2$              &  2.54  \\
$(0s)_{10}^{-2}[(1s_{1/2})(0d_{3/2})]_{10}$    &  2.34  \\
$(0s)_{10}^{-2}[(0p_{3/2})(0f_{5/2})]_{10}$    &  1.90  \\
$(0s)_{10}^{-2}[(0p_{1/2})(0p_{3/2})]_{10}$    &  1.55  \\
$(0s)_{10}^{-2}[(0d_{5/2})(0g_{7/2})]_{10}$    &  0.79  \\
$(0s)_{10}^{-2}(0d_{3/2})_{10}^2$              &  0.44  \\
\mbox{remaining part}                          &  7.96  \\
\hline
\end{tabular}
\end{center}
\end{table}

In Table \ref{tab:tensor}, we list the dominant components of the matrix element of the tensor interaction between the $0p0h$ and $2p2h$ states.
We can expand the total contribution of the tensor interaction $\langle V_T \rangle$ into two types of the matrix element
between $0p0h$ and $2p2h$ states and between $2p2h$ and $2p2h$ states of the wave function.
It is found that the former $0p0h$-$2p2h$ coupling produces $-49.13$ MeV of the tensor force matrix element,
which exhausts approximately 90\% of the total value of $-54.55$ MeV in Table~\ref{tab:benchmark}. This feature was first pointed out in ATMS\cite{akaishi86}.
In particular, three specific configurations, $(0p_{1/2})^2_{10}$, $(1s_{1/2})(0d_{3/2})_{10}$ and $(0p_{3/2})(0f_{5/2})_{10}$ for particle states,  
give large contributions in Table \ref{tab:mixing}. 
They also have large mixing probabilities in the wave function in Table \ref{tab:mixing}.
These facts denote that three configurations are essential for describing the tensor correlation in $^4$He.
It is noted that the mixing probability of each configuration in Table \ref{tab:tensor} is less than 3$\%$,
but their contributions to the tensor force matrix element $\langle V_T \rangle$ are relatively large.
This feature was also discussed in the previous paper\cite{myo06}.

\begin{table}[t]
\centering
\caption{Contributions of each $0p0h$-$2p2h$ coupling to $\langle V_T \rangle$ in MeV.}
\label{tab:tensor}
\renc{\baselinestretch}{1.20}
\begin{tabular}{c|c}\hline\hline
two particle states         &  $\langle \Phi_{0p0h}|V_T|\Phi_{2p2h}\rangle$ \\
\hline
$(0p_{1/2})^2_{10}$         &  $-$8.38  \\
$(1s_{1/2})(0d_{3/2})_{10}$ &  $-$10.99 \\
$(0p_{3/2})(0f_{5/2})_{10}$ &  $-$10.17 \\
$(0d_{5/2})(0g_{7/2})_{10}$ &  $-$5.67  \\
$(0d_{3/2})^2_{10}$         &  $-$2.62  \\
$(0p_{1/2})(0p_{3/2})_{10}$ &  $-$2.48  \\
\hline
\end{tabular}
\end{table}

\section{Conclusions}\label{sec:conclusion}
We have developed a method of calculating the nuclear ground state using the nucleon-nucleon interaction in the shell model framework.  The important features of the nucleon-nucleon interaction are the strong tensor interaction caused by the pion exchange and the strong short-range repulsion caused by the internal structure of the nucleon.  We have treated the tensor interaction in terms of the tensor-optimized shell model (TOSM) in which, in addition to the core state, we introduce two-particle two-hole ($2p2h$) states, to take into account the excitations due to the tensor interaction.  As for the short-range repulsive interaction, we have introduced the unitary correlation operator method (UCOM), in which the unitary transformation is introduced to express the short-range behavior of the relative wave function.  We have then combined these two methods to calculate the nuclear ground state, which is called TOSCOM.

In TOSCOM, we have worked out the formulation in the Gaussian basis function, and all the matrix elements are given in the Appendix.  
We obtained good convergence in the calculated results of the energy and other components of the Hamiltonian.  This denotes that short-range correlation is successfully described using UCOM, and simultaneously,
the tensor correlation is also described explicitly using TOSM with the inclusion of the $2p2h$ states with high-momentum. 
Hence, we have developed the method of describing the nuclear structure starting from the nucleon-nucleon interaction in TOSCOM.

We have carried out calculations first for $^4$He using the UCOM for all the partial waves.  We have found that the tensor correlations were largely underestimated and the binding energy of $^4$He is also somewhat underestimated.  The reason is the removal of the short-range part of the relative wave function in the optimization of the short-range correlations, where the tensor interaction needs some amount of strength.  To overcome this feature, we have newly introduced $S$-UCOM, where the unitary transformation was performed only for the $s$-wave component of the relative wave function.  It was shown that the situation was largely improved, and the numerical results are found to be very close to the rigorous calculation.  This is a very encouraging result to describe nuclei using the nucleon-nucleon interaction.  We have, however, still somewhat lack of contribution of the tensor interaction matrix element.

We consider that we have carried out the best calculation in the present framework.  We have introduced one approximation of truncating the correlated operators due to the unitary transformation up to the two-body terms.  This truncation has been shown to be good in the case of the central interaction alone by Feldmeier et al.\cite{feldmeier98}, since the short-range repulsion is of short-range.  However, the results of the tensor interaction obtained in the TOSCOM formulation seems to require more attention because of the interference with the short-range repulsion. The tensor interaction seems to require some short-range components, which are taken away by the UCOM treatment of the short-range repulsion.  We are currently investigating the effect of the three-body terms and the results will be reported in the near future.  It could also be an idea to perform further UCOM with the tensor correlation operator\cite{neff03} in the very short-range part on top of TOSCOM, which can treat the intermediate- and long-range parts of the tensor correlations.  We would like to note here that the present calculation is nearly a variational calculation in the shell model basis, and the numerical results are very encouraging for expressing all the necessary correlations in the calculated wave functions caused by the nucleon-nucleon interaction.

\section*{Acknowledgements}
We are grateful to Prof. H. Horiuchi for his continued interest and fruitful discussions on the role of the tensor interaction on nuclear structure.  
This work was supported by a Grants-in-Aid from the Japan Society for the Promotion of Science (JSPS, No. 18-8665 and 18540269)
and also by the JSPS Core-to-Core Program.
Numerical calculations were performed on the computer system at the Research Center for Nuclear Physics.

\appendix
\section{Two-Body Matrix Elements in TOSM}

\nc{\Rac}[2]	{W(#1;#2)}			
\nc{\CG}[2]	{\langle #1 | #2 \rangle} 	
\nc{\EV}[1]	{\bra #1 \ket}                  
\nc{\ME}[3]	{\bra #1|#2|#3\ket}             
\nc{\br}        {{\bf r}}                       
\nc{\Pot}       {{\mathcal V}}                  
\nc{\wtil}      {\widetilde}                    
\nc{\what}      {\widehat}                      
\nc{\bsh}[1]	{\widehat{\boldsymbol{#1}}}
\nc{\sig}       {\sigma}                        
\nc{\hr}        {{\hat{\bf r}}}                 
\nc{\al}        {\alpha}                        
\nc{\lam}       {\lambda}                       
\nc{\Gam}       {\Gamma}                        
\nc{\Nj}[9]     {{\footnotesize
                        \left\{                 
                        \begin{array}{ccc}
                        #1 & #2 & #3    \\
                        #4 & #5 & #6    \\
                        #7 & #8 & #9
                        \end{array}
                        \right\}
                }}

We write here the two-body matrix elements of the central, LS and tensor interactions in the Gaussian basis function.
We also expand the potential with the finite number of the Gaussian function. 
Hence, we need to calculate the two-body matrix elements of the potential having Gaussian form. We do not write the isospin part.
The matrix elements are calculated by transforming wave functions from $jj$ coupling scheme to $LS$ coupling scheme.
In the following, we define $L$ as the coupled orbital angular momentum of the two-particle states in the $V$-coordinate bases.

First, we define the Gaussian basis function for one nucleon state having the orbital angular momentum $l,m$
and the length parameter $b$. 
\begin{eqnarray}
  u_{lm}^a(\vc{r})
&=&     N_l(a)\, r^l\, e^{-\frac{a}{2}r^2}\, Y_{lm}(\hat{\bs{r}}),\quad
        a=\frac{1}{b^2},\quad
        N_l(a)
=       \left[  \frac{2\ a^{l+3/2} }{ \Gamma(l+3/2)}\right]^{\frac12}
\end{eqnarray}
Using this basis function, we evaluate the formulae of the matrix elements for the central, LS and tensor interactions
and further the overlap between $V$ and $T$-type bases for $s$-wave two-nucleon relative motion in the $T$-type bases.

\subsection{Central interaction}
We write the central interaction as
\begin{eqnarray}
        V^C(r)
&=&     e^{-\rho(\bs{r}_1-\bs{r}_2)^2} .
\end{eqnarray}
We also consider multiplying the factor $r_1^{n_1} r_2^{n_2}$, which is used for $1s$-wave state.
The matrix element is given as
\begin{eqnarray}
&&	\hspace*{-1.0cm}
	\bras{[u^{a_1}_{l_1},u^{a_2}_{l_2}]_L}
	 r_1^{n_1} r_2^{n_2} e^{-\rho(\bs{r}_1-\bs{r}_2)^2}
	\kets{[u^{a_3}_{l_3},u^{a_4}_{l_4}]_{L}}
	\nonumber
	\\
&=&	\frac{\sqrt{\pi}}{2} \ (-1)^{l_2+l_4-L} \
	\prod_{i=1}^4\left\{ N_{l_i}(a_i)\cdot \hat{l_i}\right\}
	\nonumber
	\\
&\times&\sum_{k=0} \Rac{l_1 l_2 l_3 l_4}{Lk}
	\CG{l_1 0 l_3 0}{k 0} \CG{l_2 0 l_4 0}{k 0}\
        {\mathcal I}_{l_1,l_2,l_3,l_4}^{k,n_1,n_2}(\rho,\beta_1,\beta_2) .
\end{eqnarray}
Here, $\Rac{l_1 l_2 l_3 l_4}{Lk}$ is Racah coefficient and ${\mathcal I}_{l_1,l_2,l_3,l_4}^{\mu,n_1,n_2}(\rho,\beta_1,\beta_2)$ is defined as
\begin{eqnarray}
&&	\hspace*{-1.0cm}
	{\mathcal I}_{l_1,l_2,l_3,l_4}^{\mu,n_1,n_2}(\rho,\beta_1,\beta_2)
        \nonumber
        \\
&=&	\frac{I_{P_1+\mu+2}(\beta_1)\cdot \rho^\mu}{\Gamma(\mu+\frac32)}\
	\sum_{n=0}^{\textstyle\frac{P_1-\mu}{2}}	
	\frac{\left(\frac{\mu-P_1}{2}\right)_n}{\left(\mu+\frac32\right)_n\cdot n!}
	\left(-\frac{\rho^2}{\beta_1}\right)^n
	I_{P_2+\mu+2+2n}(\beta_3)
\end{eqnarray}
with
\begin{eqnarray}
	P_1
&=&	l_1+l_3+n_1,
	\qquad
	P_2
~=~	l_2+l_4+n_2,
	\qquad
        \hat{l}
~\equiv~\sqrt{2l+1},
	\\
	\beta_1
&=&	\frac{a_1+a_3}{2}+\rho,
	\qquad
	\beta_2
~=~	\frac{a_2+a_4}{2}+\rho,
	\qquad
	\beta_3
~=~	\beta_2 - \frac{\rho^2}{\beta_1},
\end{eqnarray}
where
\begin{eqnarray}
	I_n(a)
&=&    \int_0^{\infty} x^n e^{-ax^2} dx
=      \frac{\Gam(\frac{n+1}{2})}{2a^{\frac{n+1}{2}}}
=	\frac1{\left[N_{\frac{n}{2}-1}(a)\right]^2} .
\end{eqnarray}
Here, $(x)_n \equiv x(x+1)(x+2) \cdots (x+n-1)=\prod_{i=0}^{n-1}(x+i)=\displaystyle \frac{\Gam(x+n)}{\Gam(x)}$, where $x>0$, and $(x)_0 = 1$.

\subsection{LS interaction}
We write the LS interaction as
\begin{eqnarray}
  V^{LS}(r)
&=&     e^{-\rho(\bs{r}_1-\bs{r}_2)^2}\, (\vc{L}\cdot \vc{S}).
\end{eqnarray}
Here
\begin{eqnarray}
        \vc{L}
&=&     \vc{r}\times \vc{p},\quad
        \vc{r}
~=~     \vc{r}_1-\vc{r}_2,\quad
        \vc{p}
~=~     \frac12 (\vc{p}_1-\vc{p}_2),\quad
        \vc{S}
~=~     \vc{s}_1+\vc{s}_2.
\end{eqnarray}
The matrix element of the LS interaction is given as
\begin{eqnarray}
&&	\hspace*{-1.0cm}
	\left\langle{\left[ [u^{a_1}_{l_1},u^{a_2}_{l_2}]_L,\chi_1\right]_J}
	\bigg|{e^{-\rho \bs{r}^2} (\vc{L}\cdot \vc{S})}\bigg|
	{\left[ [u^{a_3}_{l_3},u^{a_4}_{l_4}]_{L'},\chi_1\right]_J}\right\rangle
	\nonumber
	\\
&=&	(-1)^{L+1-J}\sqrt{6}\ \Rac{L1L'1}{J1}\,
	\ME{[u^{a_1}_{l_1},u^{a_2}_{l_2}]_L}{| e^{-\rho \bs{r}^2} \vc{L}|}{[u^{a_3}_{l_3},u^{a_4}_{l_4}]_{L'}},
\end{eqnarray}
where $\chi_1$ is the coupled wave function of two intrinsic spins with triplet state.
The reduced matrix elements including the orbital angular momentum $\vc{L}$ consist of four terms as
\begin{eqnarray}
&&	\hspace*{-1.0cm}
	\ME{[u^{a_1}_{l_1},u^{a_2}_{l_2}]_L}{| e^{-\rho \bs{r}^2} \vc{L}|}{[u^{a_3}_{l_3},u^{a_4}_{l_4}]_{L'}}
        \nonumber
  \\
&=&	\frac12\ \Bigl[
        \ME{[u^{a_1}_{l_1},u^{a_2}_{l_2}]_L}{| e^{-\rho \bs{r}^2} \vc{l}_1|}{[u^{a_3}_{l_3},u^{a_4}_{l_4}]_{L'}}
+	\ME{[u^{a_1}_{l_1},u^{a_2}_{l_2}]_L}{| e^{-\rho \bs{r}^2} \vc{l}_2|}{[u^{a_3}_{l_3},u^{a_4}_{l_4}]_{L'}}
        \nonumber
	\\
&-&	\ME{[u^{a_1}_{l_1},u^{a_2}_{l_2}]_L}{| e^{-\rho \bs{r}^2} \vc{r}_1\times\vc{p}_2|}{[u^{a_3}_{l_3},u^{a_4}_{l_4}]_{L'}}
        \nonumber
	\\
&-&	\ME{[u^{a_1}_{l_1},u^{a_2}_{l_2}]_L}{| e^{-\rho \bs{r}^2} \vc{r}_2\times\vc{p}_1|}{[u^{a_3}_{l_3},u^{a_4}_{l_4}]_{L'}}
	\Bigr] .
\end{eqnarray}
Here, the first term including $\vc{l}_1$ is
\begin{eqnarray}
&&	\hspace*{-1.0cm}
	\ME{[u^{a_1}_{l_1},u^{a_2}_{l_2}]_L}{| e^{-\rho \bs{r}^2} \vc{l}_1|}{[u^{a_3}_{l_3},u^{a_4}_{l_4}]_{L'}}
	\nonumber
	\\
&=&	\frac{\sqrt{\pi}}{2}\cdot 
	\prod_{i=1}^4 N_{l_i}(a_i)\cdot
	(-1)^{l_1-l_4-1}
	\sqrt{l_3(l_3+1)}\cdot  \hat{L}\hat{L'}\cdot
        \Rac{l_3 L l_3 L'}{l_4 1}
	\nonumber
	\\
&\times&\hat{l_1} \hat{l_2} (\hat{l_3})^2 \hat{l_4}
        \sum_{\mu}
	\CG{l_1 0 l_3 0}{\mu 0}
	\CG{l_2 0 l_4 0}{\mu 0}
	\Rac{l_1 l_2 l_3 l_4}{L \mu}
        \nonumber
        \\
&\times&{\mathcal I}_{l_1,l_2,l_3,l_4}^{\mu,0,0}(\rho,\beta_1,\beta_2),
\end{eqnarray}
and the second term including $\vc{l}_2$ is
\begin{eqnarray}
&&	\hspace*{-1.0cm}
	\ME{[u_{l_1},u_{l_2}]_L}{| e^{-\rho \bs{r}^2} \vc{l}_2|}{[u_{l_3},u_{l_4}]_{L'}}
	\nonumber
	\\
&=&	\frac{\sqrt{\pi}}{2}\cdot 
	\prod_{i=1}^4 N_{l_i}(a_i)\cdot
	(-1)^{l_1-l_4+L+L'-1}
	\sqrt{l_4(l_4+1)}\cdot \hat{L}\ \hat{L'}\cdot
        \Rac{l_4 L l_4 L'}{l_3 1}
	\nonumber
	\\
&\times&\hat{l_1} \hat{l_2} \hat{l_3} (\hat{l_4})^2
        \sum_{\mu}
	\CG{l_1 0 l_3 0}{\mu 0}	\
	\CG{l_2 0 l_4 0}{\mu 0}	\
	\Rac{l_1 l_2 l_3 l_4}{L \mu}
        \nonumber
        \\
&\times&{\mathcal I}_{l_1,l_2,l_3,l_4}^{\mu,0,0}(\rho,\beta_1,\beta_2).
\end{eqnarray}
The other reduced matrix elements are 
\begin{eqnarray}
&&	\ME{[u^{a_1}_{l_1},u^{a_2}_{l_2}]_L}{| e^{-\rho \bs{r}^2} \bs{r}_1\times\vc{p}_2\ |}{[u^{a_3}_{l_3},u^{a_4}_{l_4}]_{L'}}
         \nonumber
	\\
&=&	\frac{\sqrt{\pi}}{2}\cdot
	\prod_{i=1}^4 N_{l_i}(a_i)\cdot
	\sqrt{6}\cdot \hat{L}\ \hat{L'}\cdot
        \hat{l_1}\ \hat{l_2}\ \hat{l_3} 
        \sum_{\lambda_3,\lambda_4}\
	(-1)^{l_1-\lambda_3+L}\	\CG{l_3 0 1 0}{\lambda_3 0}
        \nonumber
	\\
&\times&
	\Nj{\lambda_3}{\lambda_4}{L}{l_3}{l_4}{L'}{1}{1}{1}\cdot
	\hat{\lambda_3}\ \hat{\lambda_4}
        \sum_{\mu}
	\CG{l_1 0 \lambda_3 0}{\mu 0}
	\CG{l_2 0 \lambda_4 0}{\mu 0}
	\Rac{l_1 l_2 \lambda_3 \lambda_4}{L \mu}
        \nonumber
	\\
&\times&\biggl[\ a_4\ {\mathcal I}_{l_1,l_2,l_3,l_4}^{\mu,1,1}(\rho,\beta_1,\beta_2)
	\left( \sqrt{l_4+1}\cdot \delta_{\lambda_4,l_4+1} - \sqrt{l_4}\cdot\delta_{\lambda_4,l_4-1}\right)
        \nonumber
        \\
&+&     \sqrt{l_4}\cdot( 2l_4+1)\ {\mathcal I}_{l_1,l_2,l_3,l_4}^{\mu,1,-1}(\rho,\beta_1,\beta_2)
        \cdot \delta_{\lambda_4,l_4 -1}
	\biggr],
\end{eqnarray}
and
\begin{eqnarray}
&&	\ME{[u^{a_1}_{l_1},u^{a_2}_{l_2}]_L}{| e^{-\rho \bs{r}^2} \bs{r}_2\times\vc{p}_1|}{[u^{a_3}_{l_3},u^{a_4}_{l_4}]_{L'}}
        \nonumber
	\\
&=&	\frac{-\sqrt{\pi}}{2}\cdot\prod_{i=1}^4 N_{l_i}(a_i)\cdot
	\sqrt{6}\cdot \hat{L}\ \hat{L'}\cdot
        \hat{l_1}\ \hat{l_2}\ \hat{l_4} 
        \sum_{\lambda_3,\lambda_4}\
	(-1)^{l_1-\lambda_3+L}\	\CG{l_4 0 1 0}{\lambda_4 0}\
        \nonumber
	\\
&\times&\Nj{\lambda_3}{\lambda_4}{L}{l_3}{l_4}{L'}{1}{1}{1}\cdot
	\hat{\lambda_3}\ \hat{\lambda_4}
        \sum_{\mu}
	\CG{l_1 0 \lambda_3 0}{\mu 0}
	\CG{l_2 0 \lambda_4 0}{\mu 0}
	\Rac{l_1 l_2 \lambda_3 \lambda_4}{L \mu}
        \nonumber
	\\
&\times&\biggl[\ a_3\ {\mathcal I}_{l_1,l_2,l_3,l_4}^{\mu,1,1}(\rho,\beta_1,\beta_2)
	\left( \sqrt{l_3+1}\cdot \delta_{\lambda_3,l_3+1} - \sqrt{l_3}\cdot\delta_{\lambda_3,l_3-1}\right)
        \nonumber
        \\
&+&	\sqrt{l_3}\cdot( 2l_3+1)\ {\mathcal I}_{l_1,l_2,l_3,l_4}^{\mu,-1,1}(\rho,\beta_1,\beta_2)
        \cdot \delta_{\lambda_3,l_3 -1}
	\biggr].
\end{eqnarray}
Here, $\{\cdots\}$ including nine numbers is $9j$ symbol.
\subsection{Tensor interaction}
We write the tensor interaction as
\begin{eqnarray}
	V^T
&=&	r^m\, e^{-\rho \bs{r}^2}\, S_{12},
\end{eqnarray}
where
\begin{eqnarray}
	S_{12}
&=&	\frac{3(\vc{\sig}_1\cdot\bs{r})(\vc{\sig}_2\cdot\bs{r})}{r^2} - \vc{\sig}_1\cdot \vc{\sig}_2
~=~	\sqrt{\frac{24\pi}{5}}\ \left([\vc{\sig}_1,\vc{\sig}_2]_2\cdot Y_2(\hat{\bs{r}})\right) .
\end{eqnarray}
The matrix element of the tensor interaction is given as
\begin{eqnarray}
&&      \hspace*{-1.0cm}
	\left\langle{\left[ [u^{a_1}_{l_1},u^{a_2}_{l_2}]_L,\chi_1\right]_J}
	\bigg|r^m\, e^{-\rho\bs{r}^2}\, S_{12}\bigg|
	{\left[ [u^{a_3}_{l_3},u^{a_4}_{l_4}]_{L'},\chi_1\right]_{J}}\right\rangle
	\nonumber
	\\
&=&	4\ \sqrt{6\pi}
	(-1)^{L+1-J}\ \Rac{L1L'1}{J2}
	\nonumber
	\\
&\times&\ME{[u_{l_1},u_{l_2}]_L}{|r^m\ e^{-\rho\bs{r}^2} Y_2(\hat{\bs{r}})|}{[u_{l_3},u_{l_4}]_{L'}},
\end{eqnarray}
where
\begin{eqnarray}
&&      \hspace*{-1.0cm}
	\ME{[u^{a_1}_{l_1},u^{a_2}_{l_2}]_L}{|r^m e^{-\rho \bs{r}^2}\ Y_2(\hat{\bs{r}})|}{[u^{a_3}_{l_3},u^{a_4}_{l_4}]_{L'}}
	\nonumber
	\\
&=&	\prod_{i=1}^4 N_{l_i}(a_i)\cdot
	\sum_{p_1,p_2,\lambda_1,L_1 \atop{q_1=l_1-p_1,\atop{q_2=l_2-p_2}}}\ 
	C^{l_1,l_2,L}_{p_1,p_2,\lambda_1,L_1}
	\sum_{p_3,p_4,\lambda_2,L_2 \atop{q_3=l_3-p_3,\atop{q_4=l_4-p_4}}}\ 
	C^{l_3,l_4,L'}_{p_3,p_4,\lambda_2,L_2}\
	\hat{L}\ \hat{L'}
	\nonumber
	\\
&\times&(-1)^{\lambda_2+L_2}\ \hat{\lambda_1}\hat{\lambda_2}\hat{L_1}\hat{L_2}\cdot
	\sum_s\CG{\lambda_1 0 \lambda_2 0 }{s 0}\ \hat{s}\
        \sum_{\mu} 
        \frac{\hat{\mu}}{4\cdot \Gamma(\mu+\frac32)}
	\nonumber
	\\
&\times&\CG{L_1 0 L_2 0 }{\mu 0}
	\CG{\mu 0 s 0 }{2 0}
	\Nj{\lambda_1}{L_1}{L}{\lambda_2}{L_2}{L'}{s}{\mu}{2}
	\nonumber
	\\
&\times&{\mathcal I}_{p_1+p_2,q_1+q_2,p_3+p_4,q_3+q_4}^{\mu,m,0}\left(\frac{\beta}{2},\alpha,\alpha_2\right) .
\end{eqnarray}
Here, $p_1+q_1=l_1$ and $p_2+q_2=l_2$, and 
\begin{eqnarray}
        C^{l_1,l_2,L}_{p_1,p_2,\lam_1,L_1}
&=&     \frac{(-1)^{p_2}}{2^{p_1+p_2}}
        \sqrt{\displaystyle\frac{(2l_1+1)!(2l_2+1)!}{(2p_1+1)!(2q_1+1)!(2p_2+1)!(2q_2+1)!}}
        \nonumber
        \\
&\times&\hat{p_1}\hat{p_2}\hat{q_1}\hat{q_2}\hat{l_1}\hat{l_2}
        \Nj{p_1}{q_1}{l_1}{p_2}{q_2}{l_2}{\lam_1}{L_1}{L}\
        \CG{p_1 0 p_2 0}{\lam_1 0}
        \CG{q_1 0 q_2 0}{L_1 0},
\end{eqnarray}
\begin{eqnarray}
        \al
&=&     \displaystyle\frac{1}{2}(+a_1+a_2+a_3+a_4),\qquad
        \beta
~=~     \displaystyle\frac{1}{2}(-a_1+a_2-a_3+a_4),
	\\
        \al_1 &=& \displaystyle\frac{\al}{4} + \rho,\qquad
        \al_2 ~=~ \al - \displaystyle\frac{\beta^2}{\al+4\rho}.
\end{eqnarray}

\subsection{Overlap between $V$-type basis and $T$-type basis}

We consider the overlap between $V$-type basis and $T$-type basis.
Here, we limit the case with $s$-wave relative motion for the $T$-type basis,
whose length parameters are $a_r$ and $a_R$ for $\bs{r}$, $\bs{R}$ of the $T$-type basis, respectively.
The overlap matrix element is given as
\begin{eqnarray}
&&      \ME{[u^{a_1}_{l_1},u^{a_2}_{l_2}]_L}{r_1^{n_1} r_2^{n_2} }{u^{a_r}_0 u^{a_R}_{L}} 
        \nonumber
        \\
&=&     N_{l_1}(a_1)\ N_{l_2}(a_2)\ N_{0}(a_r)\ N_{L}(a_R)\ 
        \sum_{q=0}^{L} C_q^L 
        \sum_{\lambda=0}^\infty \frac{\sqrt{\pi}}{2} \
        (-1)^{l_2-q} \	
	\nonumber
	\\
&\times&\Rac{l_1\ l_2\ q\ L-q}{L \lam}\ \hat{l_1}\hat{l_2}\hat{q}\hat{(L-q)}\
	\CG{l_1 0 q 0  }{\lam 0}\
        {\mathcal I}_{l_1,l_2,q,L-q}^{\lambda,n_1,n_2}\left(\frac{\beta}{2},\alpha,\alpha_2\right) ,
\end{eqnarray}
where
\begin{eqnarray}
	\alpha
&=&	a_r + \frac{a_R}{4},
        \qquad
	\beta
~=~	a_r - \frac{a_R}{4},
        \qquad
	P_1
~=~	l_1+q+n_1,
	\\
	P_2
&=&	l_2+L-q+n_2,
        \qquad
        C_q^L 
~=~	\frac{ 1 }{2^{L}}\ \sqrt{ \frac{ (2L+1)! }{ (2q+1)! \cdot (2L-2q+1)! } } .
\end{eqnarray}
The overlap with $d$-wave relative motion for $\vc{r}$ in $T$-type basis can be calculated in the same manner.


\begin{thebibliography}{99}
\bibitem{pieper01} S. C. Pieper and R. B. Wiringa, Annu. Rev. Nucl. Part. Sci. {\bf 51} (2001), 53.
\bibitem{pieper02} S. C. Pieper, K. Varga and R. B. Wiringa, \PRC{66,2002,044310}.
\bibitem{pudliner97}  B. S. Pudliner, V. R. Pandharipande, J. Carlson,  S. C. Pieper and R. B. Wiringa, \PRC{56,1997,1720}.
\bibitem{myo06} T. Myo, S. Sugimoto, K. Kato, H. Toki and K. Ikeda, \PTP{117,2007,257}.
\bibitem{myo07} T. Myo, K. Kat\=o, H. Toki and K. Ikeda, \PRC{76,2007,024305}.
\bibitem{sugimoto04}	S.~Sugimoto,~K. Ikeda and~H.~Toki,~\NPA{740,2004,77}.
\bibitem{ogawa06}	Y. Ogawa, H. Toki, S. Tamenaga, S. Sugimoto and K. Ikeda,~\PRC{73,2006,034301}.
\bibitem{feldmeier98} H. Feldmeier, T. Neff, R. Roth and J. Schnack, \NPA{632,1998,61}.
\bibitem{neff03} T. Neff and H. Feldmeier, \NPA{713,2003,227}.
\bibitem{kamada01} H. Kamada et al., \PRC{64,2001,044001}.
\bibitem{kamada92} H. Kamada and W. Gl\"oeckle, \NPA{548,1992,205}.
\bibitem{hiyama03}  E. Hiyama, Y. Kino and M. Kamimura, \PPNP{51}{2003}{223}.
\bibitem{varga95} K. Varga and Y. Suzuki, \PRC{52,1995,2885}.
\bibitem{kievsky97} A. Kevsky et al., Few-Body Syst. {\bf 22} (1997), 1.
\bibitem{carlson88} J. Carlson, \PRC{38,1988,1879}.
\bibitem{navratil99} P. Navr\'atil and B.R. Barrett, \PRC{59,1999,1906}.
\bibitem{novoselsky94} A. Novoselsky and J. Katriel, \PRA{49,1994,833}.
\bibitem{roth07} R. Roth and P. Navr\'atil, \PRL{99,2007,092501}.
\bibitem{otsuka06} T. Otsuka, T. Matsuo and D. Abe, \PRL{97,2006,162501}
\bibitem{brown06} B. A. Brown, T. Duguet, T. Otsuka, D. Abe and T. Suzuki, \PRC{74,2006,061303}.
\bibitem{aoyama06} S.~Aoyama, T.~Myo, K.~Kat\=o and K.~Ikeda, \PTP{116,2006,1}.
\bibitem{otsuka07} T. Otsuka, Y. Utsuno, M. Honma, T. Mizusaki, \PPNP{46}{2003}{155}.
\bibitem{lawson} D. H. Gloeckner and R.D. Lawson, \PLB{53,1974,313}.
\bibitem{roth06} R. Roth, P. Papakonstantinou, N. Paar, H. Hergert, T. Neff and H. Feldmeier, \PRC{73,2006,044312}.
\bibitem{akaishi86}  Y.~Akaishi,~Int. Rev. of Nucl. Phys. \textbf{4}, (1986) 259.
\end{thebibliography}
\end{document}